# Fabrication of a nanoparticle-containing 3D porous bone scaffold with proangiogenic and antibacterial properties.

Juan L. Paris[1,2], Nuria Lafuente-Gómez[1], M. Victoria Cabañas[1]*, Jesús Román[1], Juan Peña[1] and María Vallet-Regí[1,2]*.

[1] Dpto. Química en Ciencias Farmacéuticas (Unidad de Química Inorgánica y Bioinorgánica), Facultad de Farmacia, Universidad Complutense de Madrid; Instituto de Investigación Sanitaria Hospital 12 de Octubre (imas12), 28040-Madrid, Spain.

[2] Centro de Investigación Biomédica en Red de Bioingeniería, Biomateriales y Nanomedicina (CIBER-BBN), Spain.

* Corresponding authors e-mail address: vcabanas@ucm.es (M. V. Cabañas); vallet@ucm.es (María Vallet-Regí)

**Abstract**

3D porous scaffolds based on agarose and nanocrystalline apatite, two structural components that act as a temporary mineralized extracellular matrix, were prepared by the GELPOR3D method. This shaping technology allows the introduction of thermally-labile molecules within the scaffolds during the fabrication procedure. An angiogenic protein, Vascular Endothelial Growth Factor, and an antibiotic, cephalexin, loaded in mesoporous silica nanoparticles, were included to design multifunctional scaffolds for bone reconstruction. The dual release of both molecules showed a pro-angiogenic behaviour in chicken embryos grown *ex ovo*, while, at the same time providing an antibiotic local concentration capable of inhibiting *Staphylococcus aureus* bacterial growth. In this sense, different release patterns, monitored by UV-spectroscopy, could be tailored as a function of the cephalexin loading strategy. The scaffold surface was characterized by a high hydrophilicity, as determined by contact angle measurements, that facilitated the adhesion and proliferation of preosteoblastic cells.

# 1. Introduction

Bone tissue has a great regeneration capacity that enables the recovery of its structure and funcion after fracture, even without the formation of scar tissue. However, as an individual ages, this regeneration capacity is gradually compromised [1–3], and people older than 85 years are at least 10 times more likely to suffer a bone fracture than people in the 60-65 year range [1]. As a consecuence of the remarkable increase in life expectancy in recent decades, pathologies related to bone fragility entail great social and economic importance. Healthcare costs for osteoporosis in women older than 75 years are nearly 20 times greater than those for women in the range of 45 to 54 years [4]. Bone fractures also constitute an important cause of morbidity, with around half of those who survive one year after fracture becoming totally or partially dependent [5], and musculoskeletal disorders as a whole being the second cause for disability worldwide [6]. In this context, surgical replacement of bone tissue has become a common and necessary procedure (being estimated that around 7 million people lived with total hip or knee replacements in 2010 in the USA [7]), with different alternatives being employed in the clinical practice regarding the source of the bone replacement material. While autografts (employing bone tissue from the own patient) is generally recognised as the gold standard, the limited amount of bone tissue susceptible of being employed for this purpose and the invasiveness of the procedure severely restrain its use in the clinical setting. On the other hand, allografts from different patients and especially xenografts from other species present safety concerns related to the host´s immune response and possibility of infectious diseases. Hence, the need for synthetic materials capable of mimicking bone tissue as closely as possible has become a tremendously active area of research in recent years [8–14].

Bone is a complex tissue for which the spatial disposition of its components in a porous structure is critical for its performace. Therefore, these synthetic materials must be assembled in scaffolds which must fulfill many requirements related to their biocompatibility and biodegradability, a 3D interconnected hierarchical porous architecture and mechanical performance [15,16]. Many different materials have been evaluated in recent years including metals (titanium, magnesium...), ceramics (calcium phosphates, bioactive glasses...), polymers (collagen, chitosan, polycaprolactone, poylactic acid...) or in combination, composites [12,14,17–21]. Besides the mentioned requirements for the materials in terms of chemical properties and architecture, the recreation of a favourable microenvironment when repairing a bone defect requires the release of biologically-active molecules at the site for a determined period of time, while also ensuring the retention of their biological activity over this same timeframe. In this sense, scaffolds designed as platforms for the release of multiple drugs/biomolecules are considered as a key strategy for tissue engineering [22–24].

The vascularization of an implanted biomaterial not only requires a substrate having interconnected pores over minimum diameter but also the capacity to induce vessel formation. In this sense, the inclusion of pro-angiogenic factors (such as Vascular Endothelial Growth Factor, VEGF) can induce the succesful colonization of such porous materials by blood vessels, surpassing this first bottleneck of regenerative medicine [25,26]. In addittion, this growth factor has proven to enhance both endochondral ossification and intermembranous bone formation while, at the same time, increase stem cell recruitment [26].

On the other hand, bone defect-related infections (highlighting staphylococci infections) [27] constitute a severe burden for bone regeneration, and drastically reduce the quality of life of affected patients. Thus, it is required to develop bone regeneration scaffolds that can simultaneously avoid the appearance of infection. While the inclusion of antibiotic drugs within these materials has shown promising results [28–33], difficulties in achieving a proper control over drug release in these systems stands as an unresolved issue. Nanoparticles have been proposed and extensively studied to control the release of different types of drugs [34–36]. The inclusion of nanoparticles within the scaffolds is a powerful tool that ensures a targeted delivery, biomolecule preservation and enables different release kinetics that each of the molecules included require for its optimal benefitial effect [37–39]. In this context, mesoporous silica nanoparticles show promising characteristics as drug carriers, such as their high loading capacity, physicochemical stability and excellent biocompatibility [40,41]. One of the main advantages of the inclusion of these loaded nanoparticles into scaffolds is derived from the local delivery into the tissue, which enhances the therapeutic efficiency allowing dose reduction when compared to systemic administration.

The aim of this study was the fabrication of a novel scaffold tailored to improve its resistance against infection and promote its vacularization thus facilitating its final performance in bone regeneration. This task requires a biodegradable scaffold that acts as a temporary extracellular matrix where cells can migrate to and proliferate thus favouring tissue regeneration. The 3D porous scaffolds are composed of apatite and agarose, as structural components, plus cephalexin-loaded mesoporous silica nanoparticles and VEGF. The morphological characteristics, culture with preosteoblast cells, drug release behavior, antibacterial effect against *Staphylococcus aureus*, and angiogenesis behaviour in chicken embryos grown *ex ovo* were here investigated.

## 2. Materials and methods

### 2.1. Synthesis and characterization of materials.

*Mesoporous Silica Nanoparticles.* Silica nanoparticles (NP) were synthesized by a modified Stöber method from tetraethyl orthosilicate in the presence of cetyltrimethylammonium bromide as structure-directing agent [42,43]. The synthesis was performed under basic and very dilute conditions and at T= 80 ºC under stirring for 2 h, as described elsewhere [43]. The surfactant template was removed by ion exchange using an extracting solution of $NH_4NO_3$ (10 mg/mL) in ethanol (95 %). The particles were suspended in that medium and magnetically stirred at 75 ºC overnight. Subsequently, they were centrifuged and washed three times with deionized water and ethanol. The product was dried under vacuum at 25 ºC. Mesoporous nanoparticles were loaded with cephalexin by impregnation method [44,45]. Drug loading was performed by dispersing 100 mg nanoparticles in 25 mL of an aqueous solution of cephalexin (8 mg/mL), and magnetically stirring them 24 h at room temperature. After this time, loaded nanoparticles (CxNP) were centrifuged and washed with water 3 times to remove the cargo adsorbed on the external surface. Finally, the sample was dried under vacuum at 25 °C.

*Nanocrystalline Hydroxycarbonateapatite powder*. Nanocrystalline Hydroxycarbonateapatite (nHCA) with similar characteristics to the bone apatite was prepared by aqueous precipitation reaction at pH = 9.2 ($NH_4OH$ solution) and 37ºC according to the method reported elsewhere [46,47]. To summarize, 1 L of 1 M solution of $Ca(NO_3)_2 \cdot 4H_2O$ and 1 L of 0.6 M $(NH_4)_2HPO_4$ and 0.3 M $(NH_4)_2CO_3$ were prepared. Both solutions were simultaneously added into a reactor at a flow rate of 30 mL/min by using a peristaltic pump (MINIPULS® 3 from Gilson). After the addition, the solution was stirred during 10 min at 2000 rpm. The white solid precipitated was filtered and washed several times with hot water and then dried overnight by freeze-drying. Afterwards, the powder was sieved and the fraction of particles with sizes lower than 60 µm was used for the scaffold fabrication.

*Fabrication of Scaffolds.* Three-dimensional macroporous agarose/nHCA scaffolds, in a 50/50 ratio, and containing VEGF and cephalexin, were fabricated using a simple shaping process named GELPOR3D [48–50] (Fig. 1): Agarose powder (Sigma–Aldrich, Steinheim, Germany, for routine use), was suspended in deionized water (3.5% w/v) and heated to 90ºC with continuous stirring; once a translucent sol was achieved, temperature was gradually decreased to 45 ºC and then, the ceramic powder (nHCA) was added. In this step, cephalexin was also added to the suspension in two ways obtaining two series of scaffolds: (a) free in solution (*S-CxF*); (b) encapsulated in nanoparticles (S-CxNP). In both cases, the amount of cephalexin included into the scaffolds was 8 mg cephalexin/g nHCA. The angiogenic factor was incorporated by adding 11.5 µg of VEGF/g nHCA, previously dissolved in phosphate buffered saline solution (100 µg VEGF/mL PBS), to the above mentioned suspension. The amount of VEGF selected was chosen so that the final dose of VEGF per sample tested on the chick embryos was 0.5 µg, same dose as we had found in the literature for chick embryo angiogenesis experiments [51–53]. The resulting slurry was poured into a designed cubic mold that provides the 3D macroporosity (Fig. 1). After five minutes at room temperature complete consolidation of the bodies was reached. After withdrawal of the designed mold, the freshly prepared scaffolds could be easily cut and shaped.

*Characterization techniques.* The materials were analysed by X-ray diffraction (XRD) in a Philips X-Pert MPD diffractometer equipped with Cu Kα radiation. Thermogravimetry and Differential Temperature Analysis (TGA/DTA) were performed in a Perkin Elmer Pyris Diamond TG/DTA analyser, with 5 ºC/min heating ramps, from room temperature to 600ºC. Fourier Transform Infrared (FTIR) spectra were obtained in a Nicolet (Thermo Fisher Scientific) Nexus spectrometer equipped with a Smart Golden Gate ATR accessory. Surface morphology was analysed by Transmission Electron Microscopy (TEM) in a JEOL JEM 2100 instrument operated at 200 kV and by Scanning Electron Microscopy (SEM) in a JEOL JSM6335F Electron microscope equipped with an Energy-dispersive-X-ray analysis (EDS) detector. $N_2$ adsorption was carried out on a Micromeritics ASAP 2020 instrument: surface area was obtained by applying the BET method to the isotherm and the pore size distribution was determined by the BJH method from the desorption branch of the isotherm. The mesopore diameter was determined from the maximum of the pore size distribution curve. The zeta potential and hydrodynamic size of nanoparticles were measured in aqueous solution by means of a Zetasizer Nano ZS (Malvern Instruments) equipped with a 633 nm "red" laser. Static contact angle measurements were carried out on bulk samples using a contact angle measuring instrument (CAM200, KSV Instruments, Finland).

**2.2. *In vitro* studies.**

*Cephalexin release study: In vitro* release assays were performed by soaking nanoparticles or scaffolds in phosphate buffered saline solution (PBS 1x) at 37°C and pH 7.4 under mechanical oscillations. The time-dependent cephalexin release was followed by UV spectroscopy [54], using an UNICAM UV500 spectrophotometer, since the cephalexin is UV-active due to the aromatic rings. Absorbance values were taken at a wavelength of 260 nm. Cephalexin solutions at concentrations of 2.5-50 µg/mL were used as standards. Experiments were repeated three times and results were expressed as means ± standard deviations.

*Cytotoxicity assay and Cell proliferation:* Prior to cell culture experiments, the prepared scaffolds were immersed in complete culture medium (Dulbecco´s Modified Eagle´s Medium (DMEM) with 10% Fetal bovine serum (FBS), 1 % glutamine and 1 % Pen-Strep) and stabilized at 37 ºC overnight. Then, 20000 MC3T3-E1 preosteoblast cells were seeded inside the prepared scaffolds and cultured for other 3 days in an incubator at 37 ºC in a humidified atmosphere of 5 % $CO_2$. Then, the Alamar Blue cell proliferation/viability assay (Promega, Spain) was performed following the manufacturer´s instructions [55].

*In vitro bacterial growth* inhibition: Agar disk diffusion tests were used to examine the antibacterial effect of scaffolds loading with Cephalexin [19]. Prior to the bacterial test, the samples were sterilized by UV-light radiation during 10 min in both sides. To perform the inhibition test, disks of 16 x 4 mm were used, prepared from slurries with and without cephalexin as above described. The surface of solid agar in Petri dishes was inoculated with a suspension of *Staphylococcus Aureus* bacterial culture with a cell concentration of 1.3 x 10$^{10}$ bacteria/mL. After that, the samples were placed on the agar plates followed by incubation at 37 °C for 24 h. Then, the bacterial inhibition zone size was measured.

### 2.3 *Ex ovo* studies.

The chorioallantoic membrane (CAM) of chicken embryos grown *ex ovo* was employed as a model to evaluate the angiogenic potential of the samples prepared. Fertilized eggs were incubated from embryonic day (ED)-1 to ED-4 in a hatcher with rotation at 38 °C at 60 % humidity. On ED-4, the embryos were de-shelled following an established method [56] with the help of a Dremel drilling tool and transferred into sterilized weighing boats. A sterile square petri dish was used to cover the embryos, which were then transferred into a static humidified incubator at 38 °C, 60 % humidity and 0.5 % $CO_2$. Scaffolds with/without VEGF were placed on top of the CAM of chicken embryos on ED-10. On ED-14, the blood vessels around the samples were evaluated under a stereomicroscope and photographed using a 4x objective.

### 3. Results and Discussion

### 3.1. Mesoporous Silica Nanoparticles.

The main characteristics of $SiO_2$-nanoparticles synthesized to encapsulate cephalexin are collected in Figures 2 and S1 (Supporting information). The materials prepared were constituted by nanoparticles with a round-shape morphology and a size around 100 nm as it can be observed in TEM image (Fig. 2A). The obtained nitrogen adsorption isotherms were type IV, with characteristic surface areas (1026 m$^2$/g) and cylindrical parallel pores (pore diameter of 2.9 nm) for this type of NPs (Fig. 2B). The charge of the nanoparticles was of -30.4mV with a hydrodynamic particle diameter of *ca.* 140 nm (Fig. S1a). Typical XRD patterns of MCM-41 type materials were observed, with the three characteristic maxima (*100*), (*110*) and (*200*), confirming the 2D hexagonal order of the mesopores arrangement (Fig. S1b). The presence of cephalexin into the nanoparticles was confirmed by FTIR spectroscopy (Fig. S1c) by the appearance of new bands in the 1600-1700 cm$^{-1}$ region in the corresponding FTIR spectrum [57]. The incorporation of cephalexin into the pores provoked a decrease in the BET surface area (Figure 2B) of the nanoparticles (700 m$^2$/g) although the ordered mesoporous channel was not affected (Fig. S1b).

Nanoparticles loaded with cephalexin were immersed in a buffered aqueous solution (pH= 7.4) at 37 °C to study the release of the drug to the medium. Fig. 3 shows the release profile as a function of immersion time. Cephalexin release data could be fitted to a first-order kinetic model (equation 1) by using GraphPad Prism software:

$$Y = A(1-e^{-kt}) \qquad \text{Equation 1}$$

being $Y$ the percentage of cephalexin released at time $t$, $A$ the maximum amount of cephalexin released (in percentage), and $k$, the release kinetic constant [58,59]. The drug release exhibits two steps: a fast delivery (burst effect) was observed over the first 5 hours, followed by a slow release at longer times. The parameters of the kinetic fitting are shown in Figure 3. In the FTIR spectra after drug release experiments (Fig. S1c) vibrational bands attributable to cephalexin were not appreciated. Consequently, the maximum amount of drug released from the silica nanoparticles, around 2 %, must be similar to the cephalexin adsorbed into the mesopores of the nanoparticles.

### 3.2. Multifunctional agarose/apatite scaffolds.

The structural basis of these multifunctional scaffolds was designed considering the mineralized water-rich matrix and highly vascularized architecture of the bone tissue to be repaired. These scaffolds contained nHCA and agarose as the majority structural components which would contribute to bone growth. The inorganic component, nHCA, synthesized by a precipitation method, showed similar characteristics to the bone apatite [60]: XRD patterns were characteristic of a nanocrystalline apatite phase, FTIR spectrum showed bands corresponding to $PO_4^{3-}$, $CO_3^{2-}$ and $OH^-$ groups and SEM micrographs showed the nanometric size of nHCA particles (Fig. S2). These apatite crystals were embedded in an agarose matrix, a natural polysaccharide that acted as a gelling agent and with hydrogel behaviour due to the hydrophilic nature of its chains [61]. Both structural scaffold components can be easily degraded thus facilitating its progressive substitution by neo-formed tissue. In an aqueous medium the hybrid scaffold would behave like a reinforced hydrogel and could absorb water within its structure, swell without destruction, and maintain its overall structure in an environment that resembled the extracellular matrix [62,63].

Cephalexin, a β-lactamic antibiotic from the group of cephalosporins, was chosen due to its effectiveness against Gram-positive bacteria [57]. Traditionally, drugs have been introduced into scaffolds by an immersion/impregnation process being the drug adsorption dependent on, among others, the specific surface area of the scaffold material. However, taking into account that the shaping method employed to prepare the scaffolds, GELPOR3D, was developed at physiological temperatures, bioactive molecules, most of them thermally-labile, could be added during the scaffold fabrication process. The introduction of cephalexin into the agarose/nHCA suspension, either dissolved in an aqueous solution (S-CxF) or encapsulated into silica nanoparticles (S-CxNP, 20% weight of NP), seemed not to affect neither the gelation time/process nor the capacity to easily shape the samples. Moreover, the addition the VEGF to the scaffolds did not change these features either.

The scaffolds were designed with a hierarchical porosity that could contribute to new bone ingrowth. Figure 4 illustrates an image of the composite fabricated scaffold containing $SiO_2$-nanoparticles (Fig. 4A) as well as SEM images (Fig. 4B and 4C). The composite scaffold exhibited a 3D interconnected porous network, which was constituted by macropores around 800 µm, a size slightly smaller than the filaments forming the mold. Besides the creation of this network of interconnected "giant" pores, this method allowed to tailor this porosity percentage, calculated to be a 30%, as a function of the filament size and the grid pattern [49]. In addition, these scaffolds showed additional porosity which ranges between 100 and 50 nm that can be attributed to the space left between the nHCA particles. This hierarchical porosity distribution results critical, on one side, to facilitate an adequate vascularization thanks to the macropores, while, on the other the interparticular porosity ensures the permeation and fluid migration throughout the whole scaffold volume. In fact, this macroporosity facilitates cell intrusion and the production of extracellular matrix within the scaffold in ectopic implants [64]. A detailed examination at higher magnification shows the distribution of the nanoparticles (indicated by arrows in Fig. 4C) embedded within the agarose/ceramic hydrogel matrix.

Figure 5 shows the EDS elemental mapping of calcium, phosphorus and silicon in agarose/nHCA scaffolds without and with nanoparticles. These images indicate that the chemical elements were distributed homogeneously, that is, the crystals of apatite, as well as the $SiO_2$-nanoparticles, were homogeneously distributed throughout the agarose matrix.

In order to fulfill its purpose as a bone regeneration scaffold, our samples had to enable cell growth on their surface. Previous studies had shown that osteoblastic cells adhered to our agarose/nHCA scaffolds, proliferated and colonized its surface [63], and these scaffolds are capable of promoting *in vivo* osteogenesis in ectopic areas [64]. Now the aim was to study if this behavior was maintained in presence of the $SiO_2$ nanoparticles. Surface hydrophilicity is a commonly used parameter to predict the biological behavior of biomaterials, with hydrophilic surfaces being regarded as suitable for cell adhesion and proliferation [19,65]. Therefore, static contact angle measurements of the employed compositions were carried out. Non-porous disks having the same composition as the studied scaffolds were obtained by pouring the corresponding slurries into Petri dishes and then dried at 37 ºC overnight. Distilled water (3 $\mu$L) was added to the different disks by a motor-driven syringe at room temperature. Reported data were obtained by averaging the results of three measurements. The results obtained (Fig. 6A) show that all the compositions employed to prepare our samples provide a hydrophilic surface (contact angle between 50 and 63º, well below 90º). These results indicate that the surface of the prepared materials would be suitable for cell growth due to the materials presenting a water contact angle compatible with cell adhesion and proliferation. This was further confirmed by seeding MC3T3 preosteoblast cells on the produced scaffolds (with and without nanoparticles). Three days later, cell proliferation was evaluated through the Alamar Blue assay (Fig. 6B), which showed that cells could grow and proliferate on the prepared scaffolds at the same rate as the control culture (cells grown on a culture plate well). These results indicate that the materials were not toxic, and cell growth on its surface was not different from cells grown on the cell culture well. All of these results taken together show the good biological behavior of the prepared scaffolds, even those containing $SiO_2$ nanoparticles, which would enable the growth of bone cells on their surface, a critical step to allow proper bone regeneration in the area.

Considering the dual aim of designing scaffolds to enable bone repair and control infection, we have fabricated bone scaffolds loaded with an antibiotic, cephalexin, introduced during the fabrication process. Two types of hybrid scaffolds were synthesized: S-CxF, scaffolds containing the structural components and free cephalexin and S-CxNP, which contained the major components and cephalexin encapsulated into nanoparticles. The release profiles of cephalexin from these hybrid scaffolds in the buffered aqueous solution are collected in Figure 7. In both cases, the release data were fitted to a first-order kinetic model with an empirical correction factor ($\delta$) (Equation 2).

$$Y = A(1-e^{-kt})^{\delta} \qquad \text{Equation 2}$$

where $Y$, $A$ and $k$, were described above. The values for $\delta$ are comprised between 1 for materials that follow first-order kinetics, and 0, for materials that release the loaded drug in the very initial time of analysis [66]. Data showed in figure 7 indicated that the scaffolds released practically all the loaded cephalexin in the first 4 h when the drug was not encapsulated (S-CxF). When the cephalexin was incorporated into the nanoparticles (S-CxNP), although a burst release of the drug was observed also at very short times, the release kinetic constant was much lower (0.22 $h^{-1}$ *vs* 0.68 $h^{-1}$), and only around 40% of the drug was released after 24 h. This result indicates the achievement of a sustained release profile of the antibiotic cephalexin when the drug was incorporated inside mesoporous silica nanoparticles prior to their incorporation into our bone regenerative scaffolds.

Agar disk diffusion test after 24 h (Fig. 8) indicated that only S-CxF scaffolds, which were fabricated with free cephalexin, inhibited bacterial growth of *Staphylococcus Aureus* (outer diameter of the inhibition zone 22.2 mm), while the disk with nanoparticles with and without drug (S-CxNP and S-NP, respectively) failed to inhibit bacterial growth. These data indicate, on one hand that the cephalexin did not lose its activity during the scaffold fabrication process and on the other, that the drug released from the scaffold containing cephalexin-loaded nanoparticles was not enough to achieve the minimum inhibitory concentration needed to inhibit bacterial growth.

According to the results collected in figures 7 and 8, encapsulating cephalexin into NPs was a good approach to prolong antibiotic treatment, but due to its slow release kinetic, more than 24 h would be necessary to reach a therapeutic concentration in the surrounding environment. Then, a combination of both strategies, delivery of cephalexin directly from the scaffold matrix and also incorporated into the nanoparticles, could be a solution in clinical practice. Then, a scaffold containing both, cephalexin encapsulated into NP and free antibiotic, S-Cx(F+NP), (Fig. 9), was fabricated in a similar way to that described above. In this case the amount of CxNP included in the scaffold S-Cx(F+NP), was the same as for S-CxNP, and half of that amount was also included in its free form (final cephalexin amount was 12 mg of drug/g nHCA). The Agar disk diffusion test as well as the drug release profile corresponding to this strategy can be observed in figure 9. The appearance of a bacterial growth inhibition zone indicated that, unlike for S-CxNP, the minimum inhibitory concentration was achieved in this system (Fig. 9A). Therefore, the amount of drug included in S-Cx(F+NP) was enough to provide an initial concentration with an antimicrobial effect. The drug release profile (Fig. 9B), showed an increased initial release compared to S-CxNP that was later followed by a slow release at later time points. The kinetic release profile of the drug from this system was also fitted to the same model as explained above, and the increase in the initial release by the incorporation of free drug was further confirmed by the lower value of the parameter $\delta$ (0.3 *vs* 0.48), which indicated a larger burst release. The results obtained indicated that these conditions allow a shock dose in the first hours and a sustained cephalexin release afterwards.

A comparison among the different scaffolds of the cephalexin released percentage and the amount of cephalexin remaining in the system after 24 h is showed in figure 9C. Data included in this figure show a larger cephalexin released % for S-Cx(F+NP) compared to S-CxNP, and much lower than that of S-CxF. On the other hand, when the amount of drug remaining within the samples after 24 h of release was evaluated (Fig. 9C), it could be observed that while S-CxNP and S-Cx(F+NP) still had a significant amount of drug, S-CxF had completely released all of its content, and would be incapable of providing any kind of sustained concentration of antibiotic. Therefore, the results obtained indicated that our synthetic scaffolds were able to release an antibiotic in a sustained manner to fight infection.

Finally, blood vessel formation has been identified as a fundamental step in bone regeneration [67,68]. Therefore, inducing angiogenesis appears as a promising strategy to enhance bone regeneration. Our versatile composite scaffolds were co-loaded with an angiogenesis-promoting protein, VEGF, to ease the colonization of the porous scaffolds by newly-formed blood vessels. These blood vessels would supply the nutrients needed for the formation of new bone, also enabling the correct disposal of waste products. Since our bone regeneration supports were fabricated at low temperature, a controlled amount of VEGF could be introduced during the fabrication scaffold process. The biological activity of VEGF released from our scaffolds was studied using an *ex ovo* chicken embryo CAM model. The *ex ovo* CAM model enables an easy evaluation of angiogenic response when exposed to different materials, and even allows carrying out several experiments in a single embryo. Since chick embryos sacrificed before hatching are not considered to be animals, the bureaucratic burden associated to their use is greatly reduced compared to animal studies. The CAM of chick embryos undergoes a rapid evolution between days 8 and 15 of embryonic development, which provides us with a window of several days to start any experiment trying to analyze pro- or anti-angiogenic strategies. Scaffolds containing nanoparticles, without and with VEGF (0.5 µg/piece) [51–53], were placed on the CAM of chicken embryos on embryonic day ED-10 (Fig. 10). On ED-14, photographs of the surrounding vasculature were taken. The pictures (Fig. 10), showed a marked increase in the number of blood vessels (especially small capillaries) around the scaffold when VEGF was included in the scaffolds (Fig. 10 right, top). This positive result highlighted the great potential of this strategy, inducing the formation of new blood vessels to ensure nutrient support for the cells growing on it.

Taking all of our results together, we have successfully designed and fabricated a multifunctional composite scaffold with 3D connected porosity. The obtained scaffolds containing $SiO_2$ nanoparticles enabled the growth of bone cells on their surface. Different biologically active molecules, such us cephalexin and VEGF, were introduced simultaneously and homogeneously into the scaffold which composition resembles the extracellular matrix where water constitutes the main component. Besides, this environment would constitute an outstanding surrounding that contributes to the preservation of biomolecules against enzymatic degradation and immunologic neutralization *in vivo*. All of these characteristics show the tremendous potential of the materials here presented for their future use in bone regeneration.

## 4. Conclusions

A composite scaffold with 3D connected porosity constituted by agarose, apatite and mesopororus silica nanoparticles was fabricated. These scaffolds could perform multiple functions simultaneously: i) the surface of the material was suitable for the adhesion of preosteoblast cells, which could grow on the scaffolds without any toxicity related to any of the components of our constructs, ii) the antibiotic cephalexin could be released from the material in a controlled manner thanks to the combination of free and nanoparticle-encapsulated drug, providing local concentrations capable of inhibiting bacterial growth, and iii) the release of co-loaded VEGF was also capable of inducing the formation of new blood vessels in the area, a key characteristic to facilitate bone regeneration.

While in this work we have employed a combination of non-encapsulated and encapsulated antibiotic as a means to obtain a sustained release system with enough initial inhibitory effect, future research could try to modify the surface of the nanoparticles to enable loading larger amounts of drug, or provide the system with release-triggering mechanisms upon certain environmental changes.

## Acknowledgements

The authors would like to thank Ministerio de Economía y Competitividad, Spain (Project MAT2015-64831-R) and Instituto de Salud Carlos III, Spain (PI 15/00978) for supporting this work. Funding from the European Research Council through the Advanced Grant VERDI (ERC-2015 AdG no.694160) is gratefully acknowledged.

**FIGURE CAPTIONS**

Figure 1. Schematic representation of the fabrication method of multifunctional scaffolds for bone regeneration.

Figure 2. TEM micrographs of $SiO_2$-nanoparticles (A) and $N_2$ adsorption isotherms of nanoparticles without/with cephalexin (B).

Figure 3. Cephalexin released from nanoparticles *versus* time. The parameters of the kinetic fitting are showed on the curve.

Figure 4. Photograph corresponding to an agarose/nHCA scaffold containing $SiO_2$-nanoparticles (A); SEM micrographs of these scaffolds at different magnifications (B and C).

Figure 5. SEM images and the corresponding EDS elemental analysis corresponding to agarose/nCHA scaffolds without (left) and with (right) silica nanoparticles.

Figure 6. Micrographs showing water drop profiles on different material surfaces. Contact angles were determined as indicated (A). MCT3T3-E1 cell proliferation (measured by Alamar Blue assay) at day 3 of culture (B).

Figure 7. Cephalexin released from different scaffolds *versus* time. The parameters of the kinetic fitting are showed on the corresponding curves.

Figure 8. Agar disk diffusion test of the different materials prepared: agarose/nHCA with nanoparticles (S-NP), with free cephalexin (S-CxF) or with encapsulated cephalexin (S-CxNP).

Figure 9. S-Cx(F+NP) scaffolds: Agar disk diffusion test (A); Cephalexin release versus time (B); A comparison among the different scaffolds of the cephalexin released percentage and the amount of cephalexin remaining in the system after 24 h (C).

Figure 10. Photograph of a chick embryo with two scaffolds placed on top of the CAM on day 10 of embryonic development (left) and stereomicroscope image of the effect of the scaffolds on surrounding vasculature at ED-14, either without VEGF (right, bottom) or with VEGF loaded in the scaffolds (right, top).

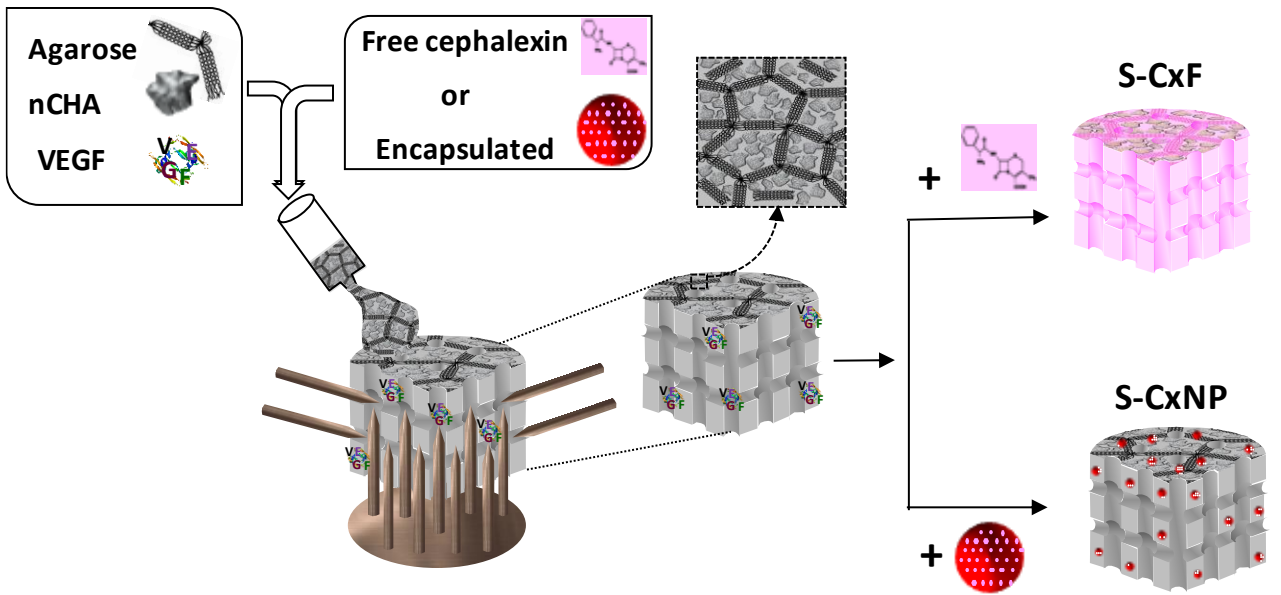

Figure 1

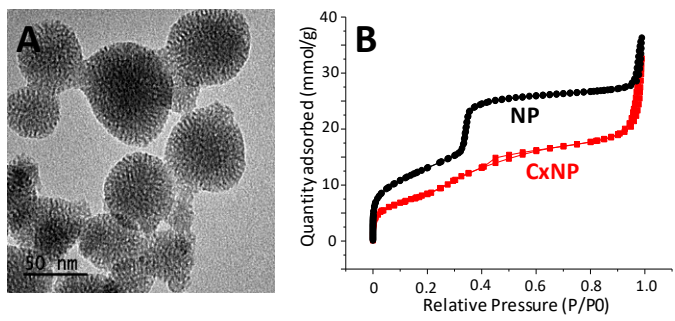

Figure 2

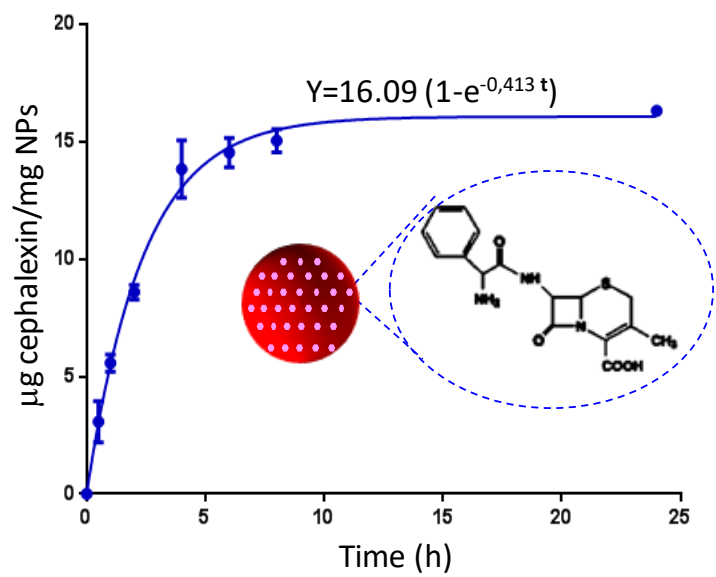

Figure 3

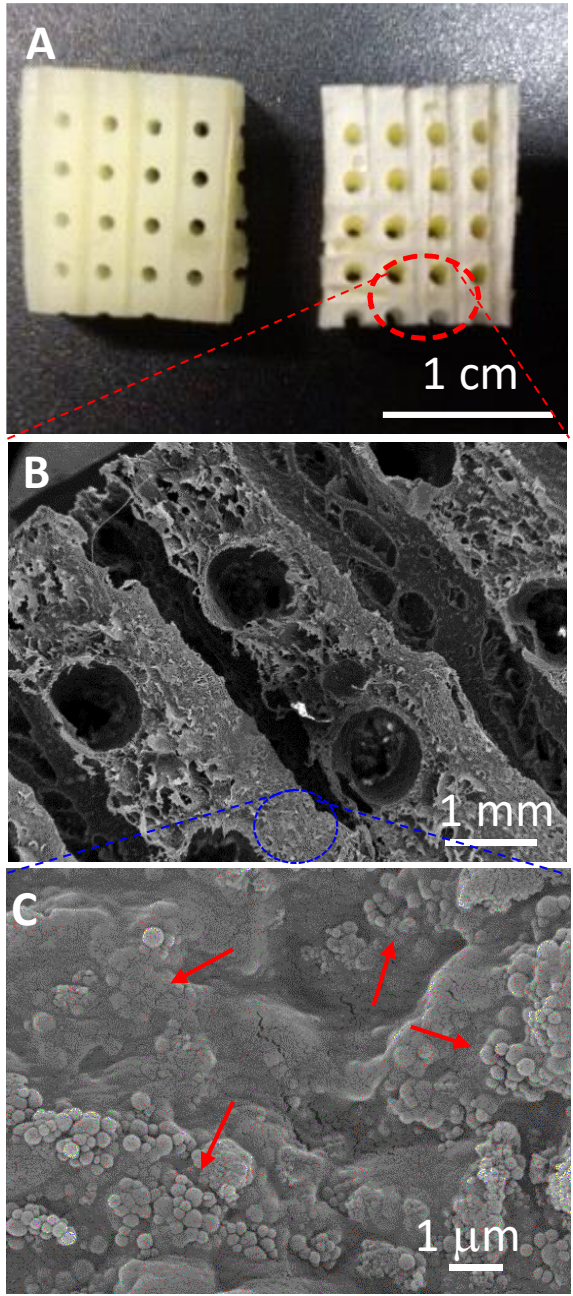

Figure 4

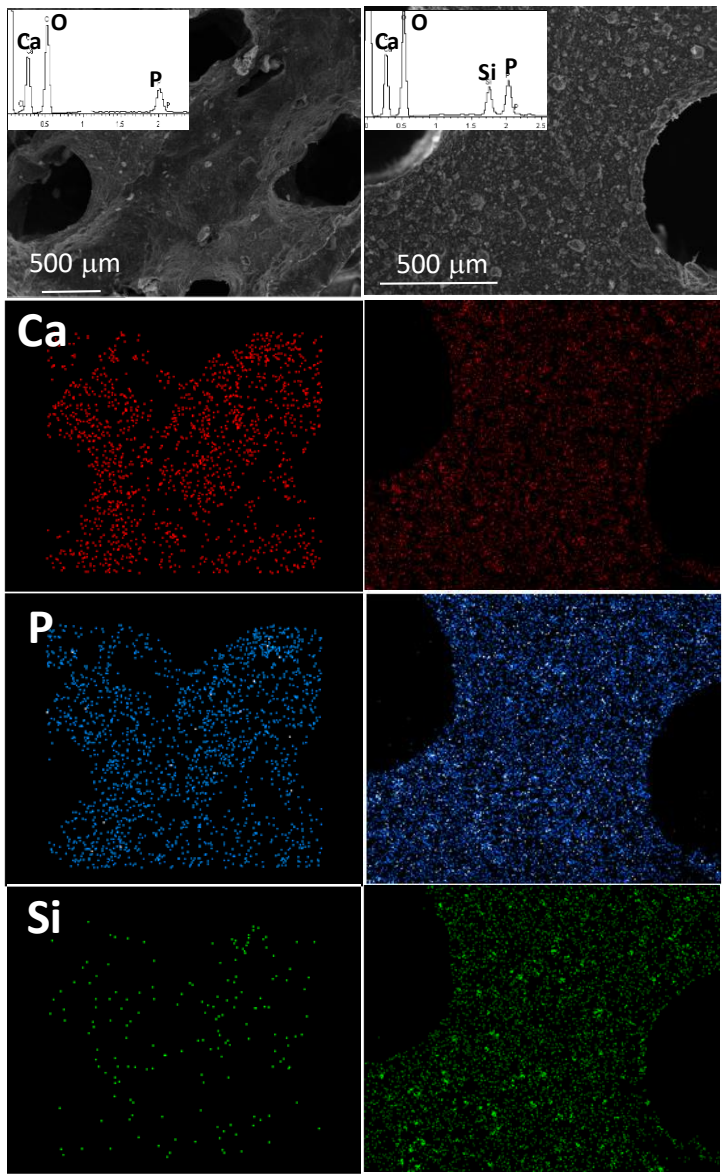

Figure 5

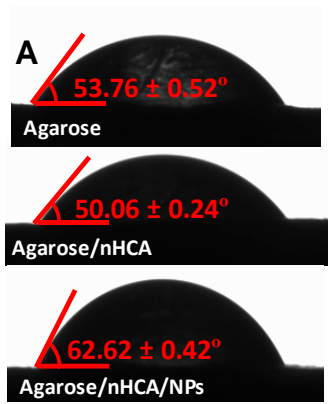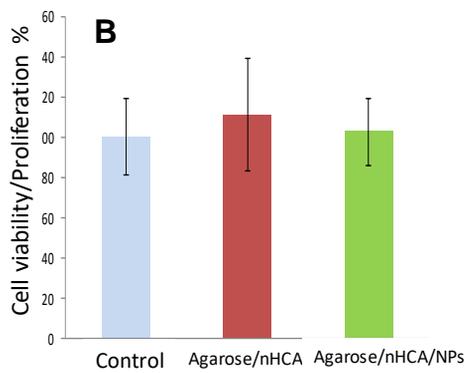

Figure 6

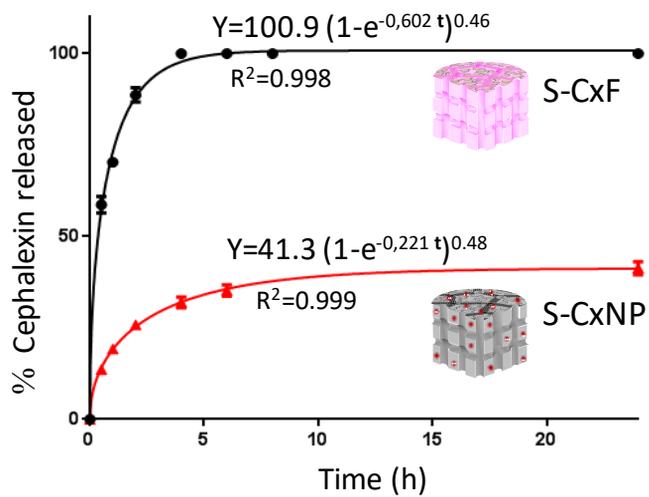

Figure 7

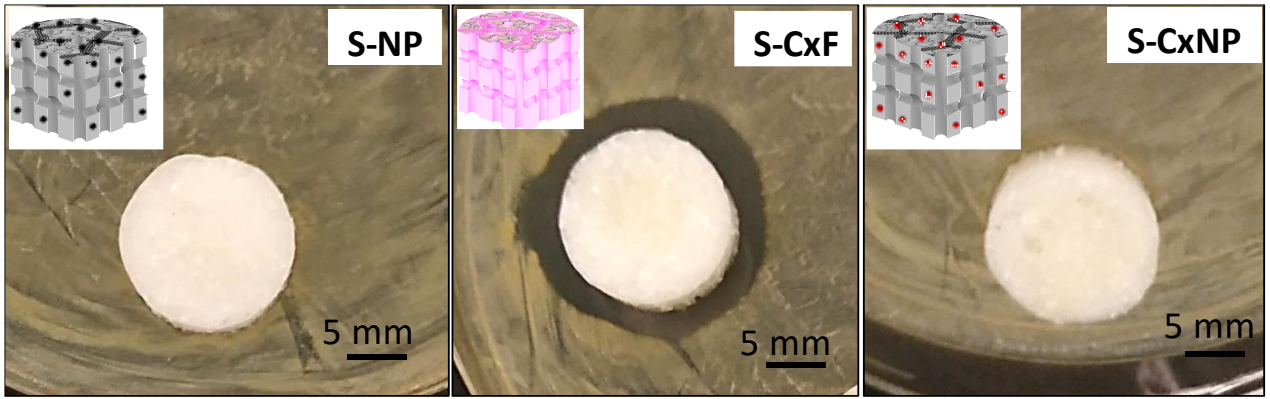

Figure 8

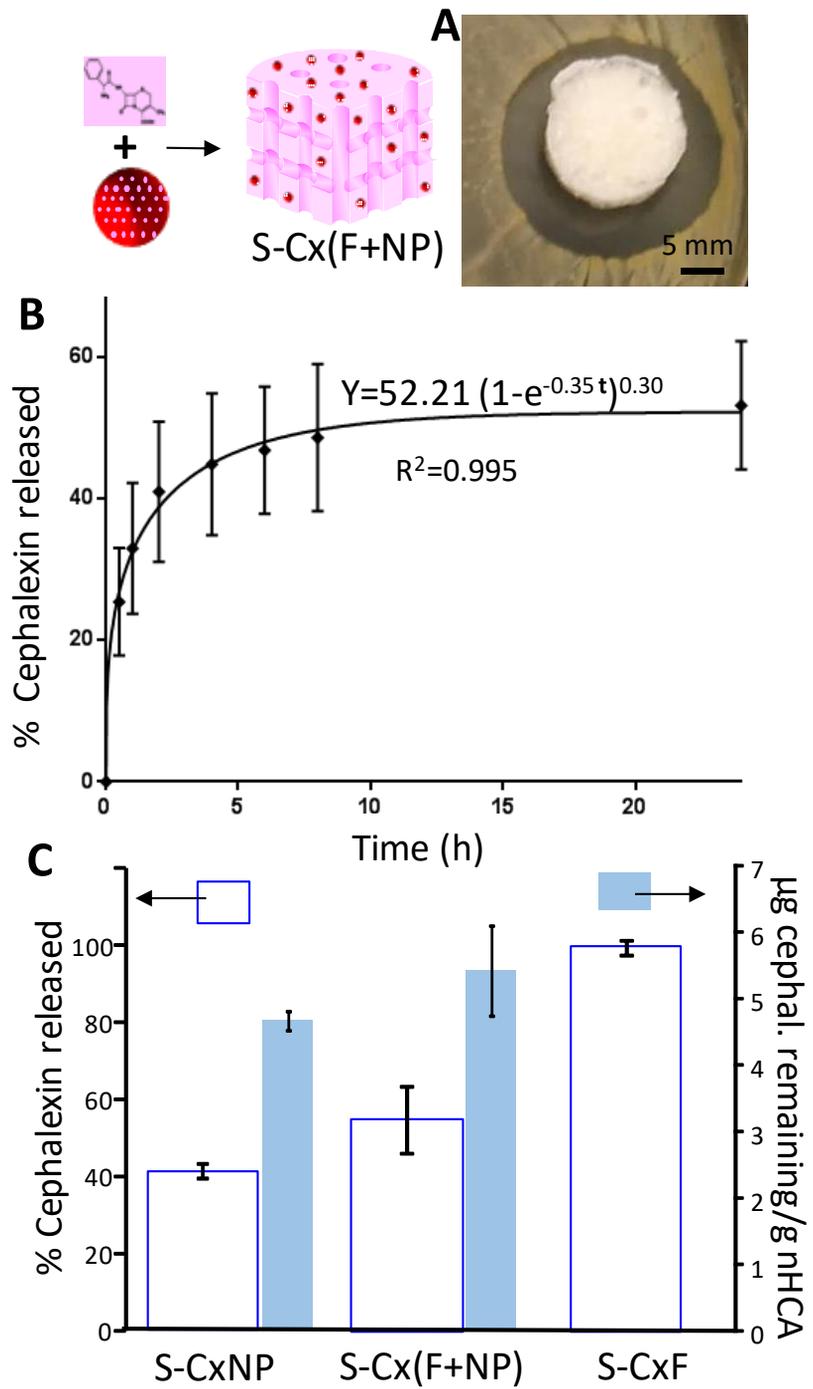

Figure 9

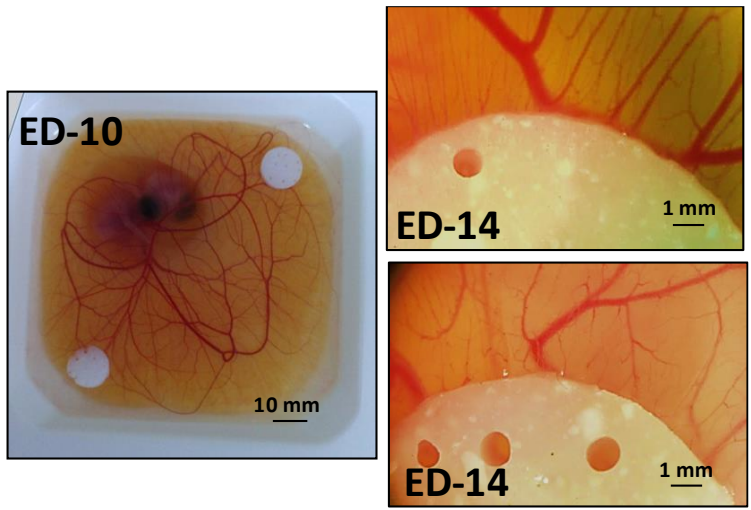

Figure 10